\documentclass[%
 aip,
 amsmath,amssymb,
 reprint,m%
]{revtex4-1}

\usepackage{graphicx}
\usepackage{dcolumn}
\usepackage{bm}

\usepackage[utf8]{inputenc}
\usepackage[T1]{fontenc}
\usepackage{mathptmx}
\usepackage{etoolbox}
\usepackage{xcolor}
\usepackage{fontawesome}
\usepackage{floatrow}	

\definecolor{linkblue}{RGB}{31,119,180}

\makeatletter
\def\@email#1#2{%
 \endgroup
 \patchcmd{\titleblock@produce}
  {\frontmatter@RRAPformat}
  {\frontmatter@RRAPformat{\produce@RRAP{*#1\href{mailto:#2}{#2}}}\frontmatter@RRAPformat}
  {}{}
}
\makeatother
\begin{document}

\preprint{AIP/123-QED}

\title[Macroscopic quantum mechanics in gravitational-wave observatories and beyond]{Macroscopic quantum mechanics in gravitational-wave observatories and beyond}
 
\author{Roman Schnabel}
\author{Mikhail Korobko}
 \email{roman.schnabel@uni-hamburg.de}
\affiliation{Institut f\"ur Laserphysik \& Zentrum f\"ur Optische Quantentechnologien, Universit\"at Hamburg,
Luruper Chaussee 149, 22761 Hamburg, Germany 
}

\date{\today}

\begin{abstract}
The existence of quantum correlations affects both microscopic and macroscopic systems. On macroscopic systems they are difficult to observe and usually irrelevant for the system's evolution due to the frequent energy exchange with the environment. 
The world-wide network of gravitational-wave (GW) observatories exploits optical as well as mechanical systems that are highly macroscopic and largely decoupled from the environment. The quasi-monochromatic light fields in the kilometre-scale arm resonators have photon excitation numbers larger than $10^{19}$, 
and the mirrors that are quasi-free falling in propagation direction of the light fields have masses of around 40\,kg.
Recent observations on the GW observatories LIGO and Virgo clearly showed that the quantum uncertainty of one system affected the uncertainty of the other.
Here, we review these observations and provide links to research goals targeted with mesoscopic optomechanical systems in other fields of fundamental physical research. These may have Gaussian quantum uncertainties as the ones in GW observatories or even non-Gaussian ones, such as Schr{\"o}dinger cat states. 
\end{abstract}

\maketitle

\section{Introduction}\label{sec:introduction}
Quantum physical experiments regarding the motion of macroscopic or even heavy bodies require low-noise and highly efficient sensing.
An ideal system is a highly reflecting mirror whose motion is sensed by monochromatic light which is photo-electrically detected with high quantum efficiency. 
The motion of the mirror needs to be isolated from any forces of the environment over the duration of the experiment. A feasible approach is a mirror suspension with a high quality factor (Q-factor). 
The suspension turns the mirror motion into that of a (quantum) mechanical oscillator. 
The higher the Q-factor the lower the coupling rate to the environment. 
A quantum optomechanical experiment is achieved if the quantum uncertainties of light and mirror motion influence each other, ultimately leading to the observation of entanglement between optical and motional degrees of freedom.

The existence of quantum uncertainties reveals itself by so-called ensemble measurements. 
These are large numbers of identical and precise measurements of the same observable performed on \emph{identical} physical systems being in \emph{identical} quantum states. 
One might assume that the preparation of an ensemble of truly identical ensemble members is impossible because of some remaining distinguishability, potentially given with respect to `hidden variables'. 
But this assumption was proven wrong through experimental violations of Bell inequalities, see for instance Ref.\cite{Aspect1981}.
The term `quantum uncertainty' describes the fact that provably identical measurement settings provide nevertheless different measurement outcomes. 
As a direct consequence of the initial indistinguishability, the individual outcomes have a truly random character, and just the outcomes' probability distribution is determined.

As given by the shape of the probability distributions, continuous-variable quantum uncertainties, such as those of position and momentum, can be either Gaussian or non-Gaussian.
The most important example of a (pure) Gaussian quantum state is the ground state. 
The most famous (pure) non-Gaussian quantum state with macroscopic excitation energy is the Schr{\"o}dinger cat state \cite{Schroedinger1935a}. 
It is a superposition of two macroscopically distinct states. 
The measurement results on an ensemble of such states are discrete and two-valued (`dead' and `alive'). 
Nevertheless, non-Gaussian states can also be represented by continuous spectra. 
Generally, the terms `quantum uncertainty' and `superposition' refer to the same physical phenomenon.
In contrast to non-Gaussian measurement spectra, Gaussian measurement spectra can generally not be used for violating a Bell inequality.
Nevertheless, some of them are as distinct as the pure non-Gaussian states. 
These are the squeezed states \cite{Walls1983,Schnabel2017}, two-mode squeezed states \cite{Ou1992}, and other related ones \cite{Bowen2002a}.
The state is called `nonclassical' if its Glauber-Sudarshan P-function does not correspond to a positive-valued (classical) probability distribution in the phase space spanned by two non-commuting observables.
The Gaussian or non-Gaussian shape of a quantum uncertainty is not relevant.

\begin{figure}
  \includegraphics[width=0.84\columnwidth]{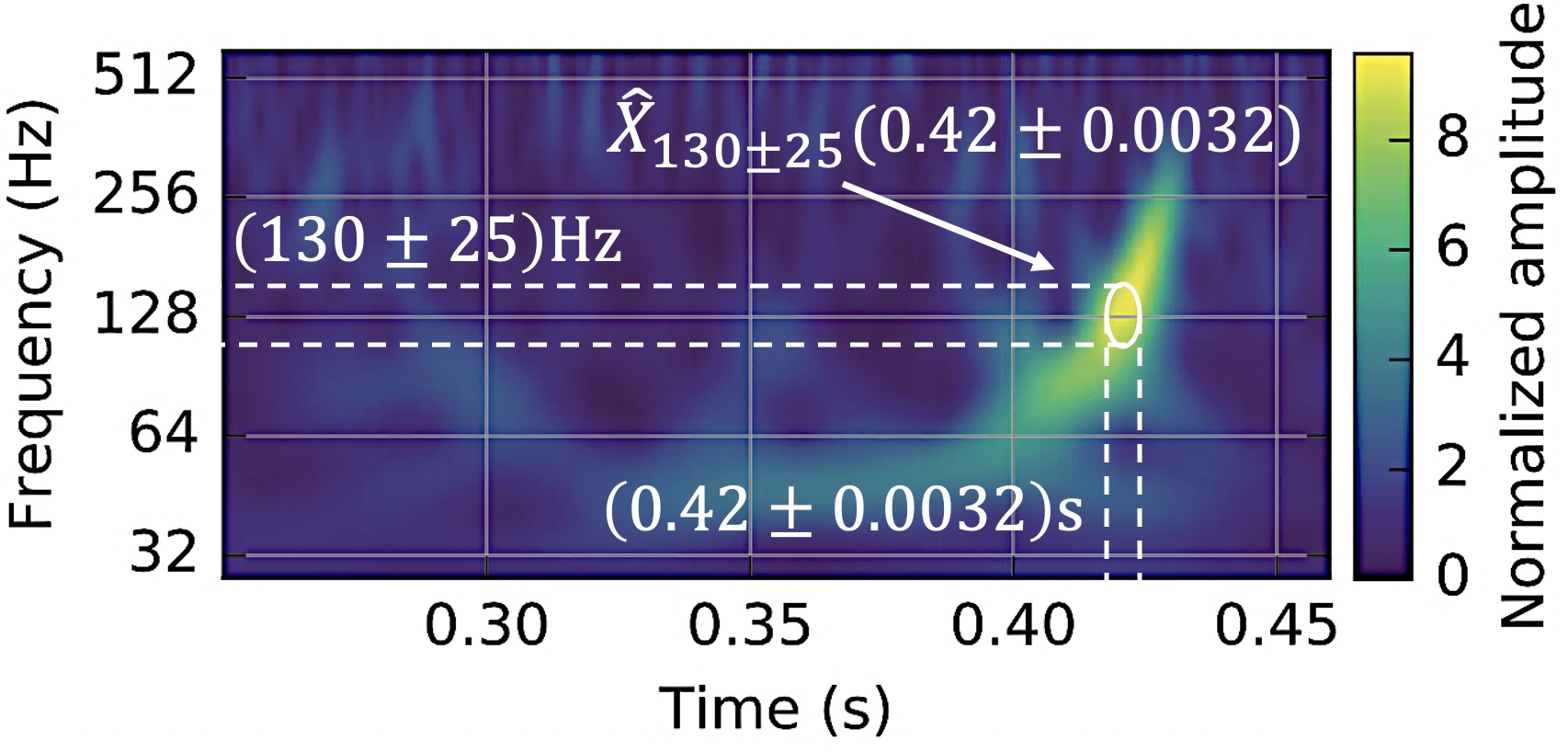}
  \vspace{-2mm}
  \caption{Time-frequency representation of the first gravitational-wave signal \cite{GW150914}. Shown are the time dependent excitations (colour coded) of the amplitude quadrature amplitudes $\hat X_{f,\Delta f}(t \pm \Delta t)$ of the observatory output light. Due to the mathematics of Fourier transform, the highlighted example $\hat X_{(130\pm25)\,{\rm Hz}}(0.42\,{\rm s}\pm0.0032\,{\rm s})$ refers to an area that has a well-defined smallest size. Reproduced with permission from Phys. Rev. Lett. 116, 061102 (2015) under CC BY \cite{GW150914}, with additions to the original to show the observable Fourier-limited mode.
  }
  \label{fig:1}
\end{figure}

\begin{figure*}
  \includegraphics[width=17cm]{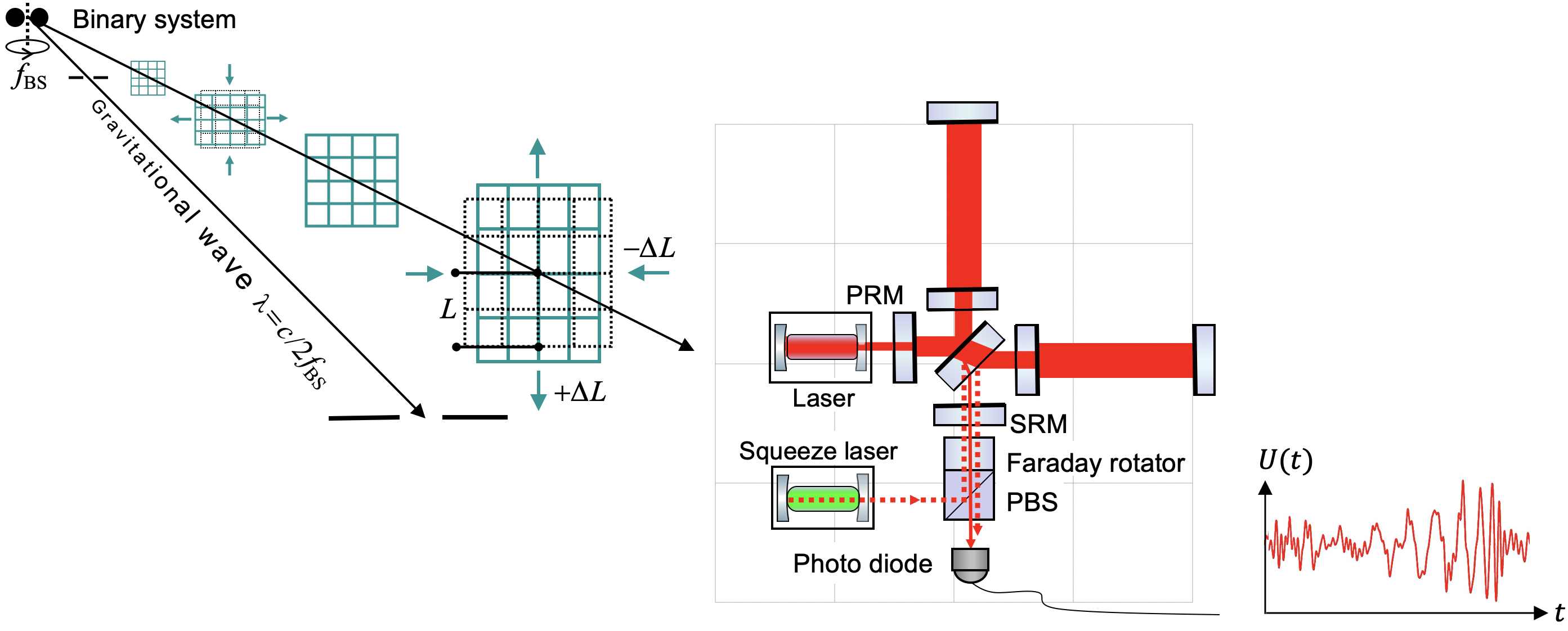}
  \vspace{0mm}
  \caption{Gravitational waves (GWs) are transverse quadrupole waves that propagate at the speed of light and expand and stretch space-time ($\pm \Delta L/L$). Sources are for instance compact binary systems of black holes and neutron stars. The LIGO and Virgo observatories are Michelson laser interferometer with an ultra-stable single mode input laser beam of about 100\,W, km-scale arm resonators, power- and signal recycling resonators, a squeeze laser for the targeted signal spectrum from 10\,Hz to 10\,kHz, and a high-quantum-efficiency PIN photo diode in the output port. The time-frequency spectrum of the photo-electric voltage resembles the GW signal plus observatory noise. $2f_{BS}$: Frequency of the GW; PRM: power-recycling mirror; SRM: signal-recycling mirror; PBS: polarising beam splitter; $U(t)$: photo-electric voltage with AC-signal.}
  \label{fig:2}
\end{figure*}

Squeezed states for improving interferometric GW observatories were proposed in 1981 \cite{Caves1981}, two years after they were proposed for bar detectors  \cite{Hollenhorst1979}. 
Squeezed states of light were first observed in the mid 1980s \cite{Slusher1985,Wu1986} and in the following years their parameters optimised for the stable and efficient exploitation in GW observatories \cite{McKenzie2004,Vahlbruch2005,Vahlbruch2006,Vahlbruch2007,Vahlbruch2008,Vahlbruch2010,Schnabel2010}.
Since 2010, the gravitational-wave detector GEO\,600 has been using (Gaussian) squeezed states of light to improve its sensitivity \cite{LSC2011,Grote2013,Lough2021}. 
Since 2019, also LIGO and Virgo have been using squeezed states of light \cite{Tse2019,Acernese2019} to improve actual observations \cite{GWTC2021}.
In the case of LIGO and Virgo, recent observations clearly showed that the quantum uncertainty of the light affected the uncertainty of the mirror motion \cite{Yu2020, Acernese2020}.
This occurs due to the optomechanical coupling, where the uncertainty in the light amplitude couples to the motion of the mirror through radiation pressure.
The phase uncertainty of the reflected light then increases, which can subsequently be detected.

Here, we review the recent observations in LIGO and Virgo that proved that the quantum uncertainties of the light fields in the arm resonators are coupled to the motion of the pendulum suspended test masses of space-time. 
The coupled, quantum-correlated uncertainties are Gaussian but nevertheless can in principle be used to set upper bounds to `spontaneous decoherence', i.e.~to hypothetical localisations of wave function due to interaction of masses with space-time or with some unknown stochastic process.
Experimental tests of gravity decoherence on non-Gaussian states of micro-mechanical oscillators have been considered almost 20 years ago \cite{Marshall2003}. 
We conclude this review by contrasting the Gaussian nonclassical states of macroscopic optomechanical systems in GW observatories with the non-Gaussian states of mesoscopic optomechanical systems. \\

\emph{The spectrum of observables of GW observatories --} 
GW observatories are Michelson-type laser interferometers with Fabry-Perot arm resonators, see Fig.~\ref{fig:2}. 
The latter are established by two pendulum-suspended laser mirrors each, separated by three kilometres in Virgo \cite{Acernese2019} and four kilometres in LIGO \cite{Tse2019}, respectively. 
During observation runs, GW observatories are in continuous operation with steady state Gaussian quantum uncertainties. During the detection of a GW or in case of disturbances, the quantum observables of related frequencies $f$ show uncertainties that are displaced over some finite time period.

Gravitational waves modulate the arm length of the interferometric GW observatory, and the signal appears at the output as the amplitude modulation of the light field, with targeted signal frequencies $f$ range from about 10\,Hz to a few kHz.
Fig.\,\ref{fig:1} shows the famous signal of the first ever measured signal (GW150914)~\cite{GW150914}. 
The signal was produced by two merging black holes and shows a frequency chirp (exponential increase with time). 
The signal looks fuzzy, because a single frequency at a single point of time is unphysical.
The relevant quantum observable in the detector is the amplitude of the amplitude quadrature of the output laser beam, $\hat X_{f,\Delta f}(t \pm \Delta t)$, defined in some frequency band $\Delta f$ around the Fourier frequency $f$ and time span $\Delta t$ at time $t$.

\begin{figure}
  \includegraphics[width=0.8\columnwidth]{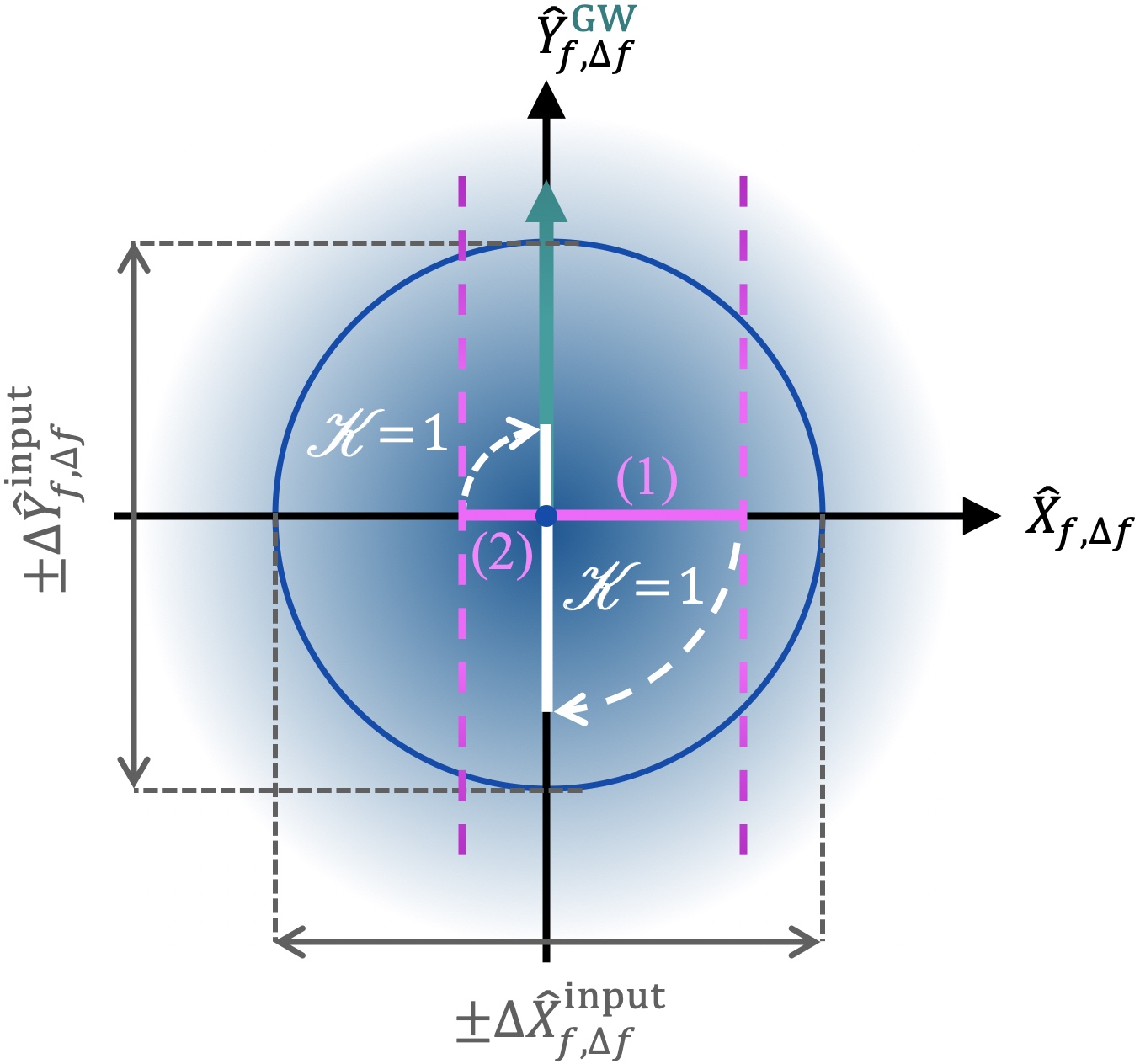}
  \vspace{-1mm}
  \caption{Illustration of quantum back action --  The fussy circular area represents the quantum uncertainty of the ground state located at the origin of the quadrature phase space. Here, the ground state is that of the input light's modulation mode with frequency $f = f_{\rm SQL}$, at which the optomechanical coupling factor $\mathcal{K}(f_{\rm SQL}) = 1$. The two vertical dashed (magenta) lines represent two example values (1) and (2) of the light's amplitude quadrature $\hat X$. The \emph{positive} value (1) produces a larger radiation pressure than the average one, and the resulting mirror movement \emph{lengthens} the optical path length. Consequently, and because of $\mathcal{K}(f_{\rm SQL}) = 1$, the phase quadrature $\hat Y$ is delayed by the same value (1). This coupling is illustrated by the large curved dashed arrow. Any \emph{negative} value (2) \emph{advances} the phase quadrature. The quantum uncertainty of the reflected light in $\hat Y$ thus corresponds to that of two uncorrelated units of ground state uncertainty. The displacement due to a gravitational wave (arrow pointing along $\hat Y$), which is a classical modulation at $f_{\rm SQL}$, is unchanged but measured with a halved signal to noise ratio.
}
  \label{fig:3}
\end{figure}

\begin{figure}
  \includegraphics[width=0.8\columnwidth]{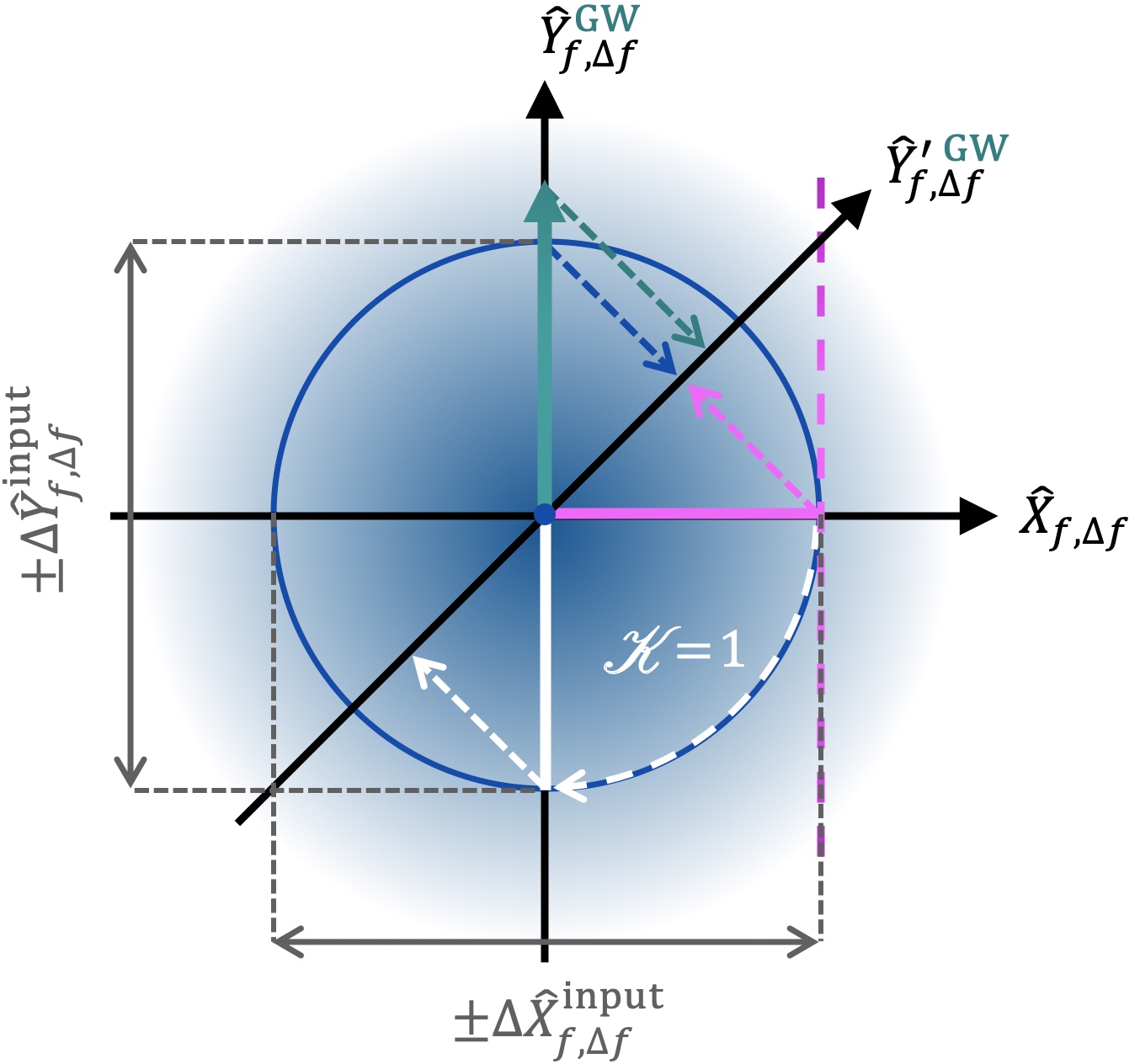}
  \vspace{-1mm}
  \caption{Illustration of back action evasion --  The situation resembles the one in Fig.\,\ref{fig:3}, but here the photoelectric detection measures the balanced linear combination of amplitude and phase quadrature modulations, i.e.~$\hat Y'$. (This is possible with a balanced homodyne detector or by rotating the phase space with a detuned filter cavity in the output port \cite{Kimble2001}. Neither option is realised in current GW observatories.) The upper two dashed arrows illustrate that the GW signal as well as the quantum measurement noise are scaled down by projecting them onto the new phase space direction. At SQL, where $\mathcal{K}(f_{\rm SQL}) = 1$, they are reduced by the factor $1/\sqrt{2}$. The important fact is the full cancellation of quantum back action if the quadrature $\hat Y'$ is detected. This is illustrated by the two lower dashed arrows, whose projections cancel each other. The overall signal to noise ration corresponds to that of zero quantum back action. Current GW observatories, however, can only read out the quadrature $\hat Y$.
}
  \label{fig:4}
\end{figure}

Observables are always defined with respect to a physical system. 
Here, these systems are light fields as well as oscillations of the mirror motion. 
A very suitable name for any of these physical systems is the word `mode'. 
A `mode' should be defined as being Fourier limited. 
An ensemble of identical such modes might be in a pure quantum state or in a rather mixed thermal state.
The observable $\hat X_{f,\Delta f}(t \pm \Delta t)$, however, addresses a \emph{specific, single} Fourier-limited mode of modulation of the monochromatic carrier output light of the GW observatory, see Fig.~\ref{fig:1}. 
The mode has the eigenfrequency $f \pm \Delta f$, when its energy is absorbed over the time period $t \pm \Delta t$.
This is mathematically described by the Fourier transform according to which the smallest phase space area for an \emph{energy} distribution is $\Delta f \cdot \Delta t = 1/(4\pi)$, also providing the well-known energy-time uncertainty relation, see e.g.~Ref.\,\cite{Schnabel2020}.
An ensemble of this mode with the same excitation is not available since the GW does not repeat itself. 
An example of the excitation of single modulation modes due to a GW is shown in Fig.~\ref{fig:1}.
The important fact is that the displayed `modes' are defined by the data post-processing. 
At first instance, the photo-electric voltage is sampled at a rate that is significantly higher than the highest expected frequency component of the GW signal.
Only after the content of frequency components has been analysed the modes of half widths $\Delta f$ and $\Delta t$ are defined.
The much higher eigenfrequency of the optical carrier field is not relevant.

The relevant mechanical system in LIGO and Virgo is composed of four mirrors, the `test masses of space-time'. They are suspended as pendulums with eigen (resonance) frequencies slightly below 1\,Hz.
This frequency is below the targeted signal spectrum. 
In order to understand how the motions of the mirrors affect the sensitivity of the GW observatory, we need to define Fourier limited `overtone' modes of the pendulum. 
The relevant modes of motion have the position excitation $\hat x_{f,\Delta f}(t \pm \Delta t)$ and the momentum excitation $\hat p_{f,\Delta f}(t \pm \Delta t)$. 
In GW observatories, these operators are one-dimensional along the laser beams' optical axes, and the position and momentum observables that actually couple to $\hat X_{f,\Delta f}(t \pm \Delta t)$ describe the difference of the two arms and are the corresponding combination of the four mirrors' individual quantities.

\emph{Quantum optomechanics in mesoscopic systems --} 
Quantum effects have also been observed in small-scale devices, where the mechanical oscillators have typical masses in the nanogram and microgram regime.
The optomechanical coupling strength is stronger for smaller masses, and for the same optical power the uncertainty of the reflected light's radiation pressure has a much stronger effect in small-scale systems.
In case of continuous steady state measurements, the quantum uncertainties in such systems are also Gaussian.
In the past years, motional ground state across all scales was achieved (see~\cite{Whittle2021} for overview), quantum radiation pressure was observed~\cite{Rossi2019,Aggarwal2020} and evaded~\cite{Moeller2017,Yap2020a, Cripe2020, Mason2019} in various systems, and Gaussian entanglement was observed~\cite{Ockeloen-Korppi2018, Chen2020a,Thomas2021,Kotler2021}.
Other systems were observed in non-Gaussian quantum states~\cite{Riedinger2016,Hong2017,Marinkovic2018}.
There, however, the relevant observable was not the quadrature, but the photon number and correlations between the photons.
So-far, no non-Gaussian state tomography on the optomechanical system has been demonstrated.
A more complete overview of recent advances in quantum optomechanics can be found in Ref.~\cite{Carney2021}, and of the progress in a rising field of levitated optomechanics in Ref.~\cite{Gonzalez-Ballestero2021}.

\begin{figure}
  \includegraphics[width=0.8\columnwidth]{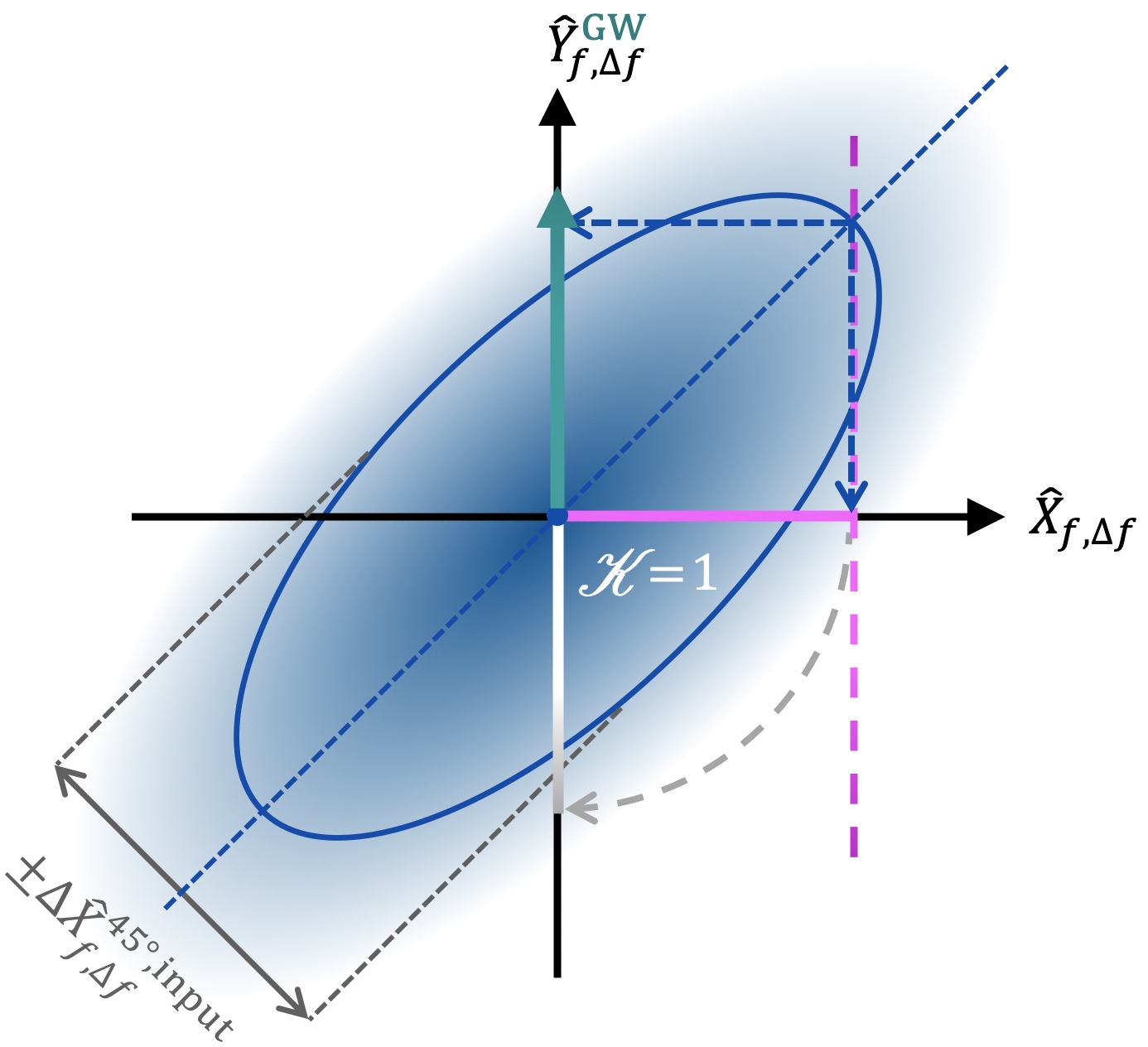}
  \vspace{-1mm}
  \caption{Back action evasion in $\hat Y$ --  The fuzzy ellipse with squeeze angle $\theta = 45^\circ$ represents the injected state at $f_{\rm SQL}$, where $\mathcal{K}(f_{\rm SQL}) = 1$. Shown is that the radiation pressure that drives $\hat X$ has a dominating \emph{correlated} component in the uncertainty of $\hat Y$. The correlation is the stronger the larger the squeeze parameter is. The radiation pressure back action produces an anti-correlated component in the uncertainty of $\hat Y$ (curved dashed arrow). The two components cancel in the conventional readout quadrature angle, which is aligned to the GW signal.
}
  \label{fig:5}
\end{figure}

\begin{figure}
  \includegraphics[width=0.8\columnwidth]{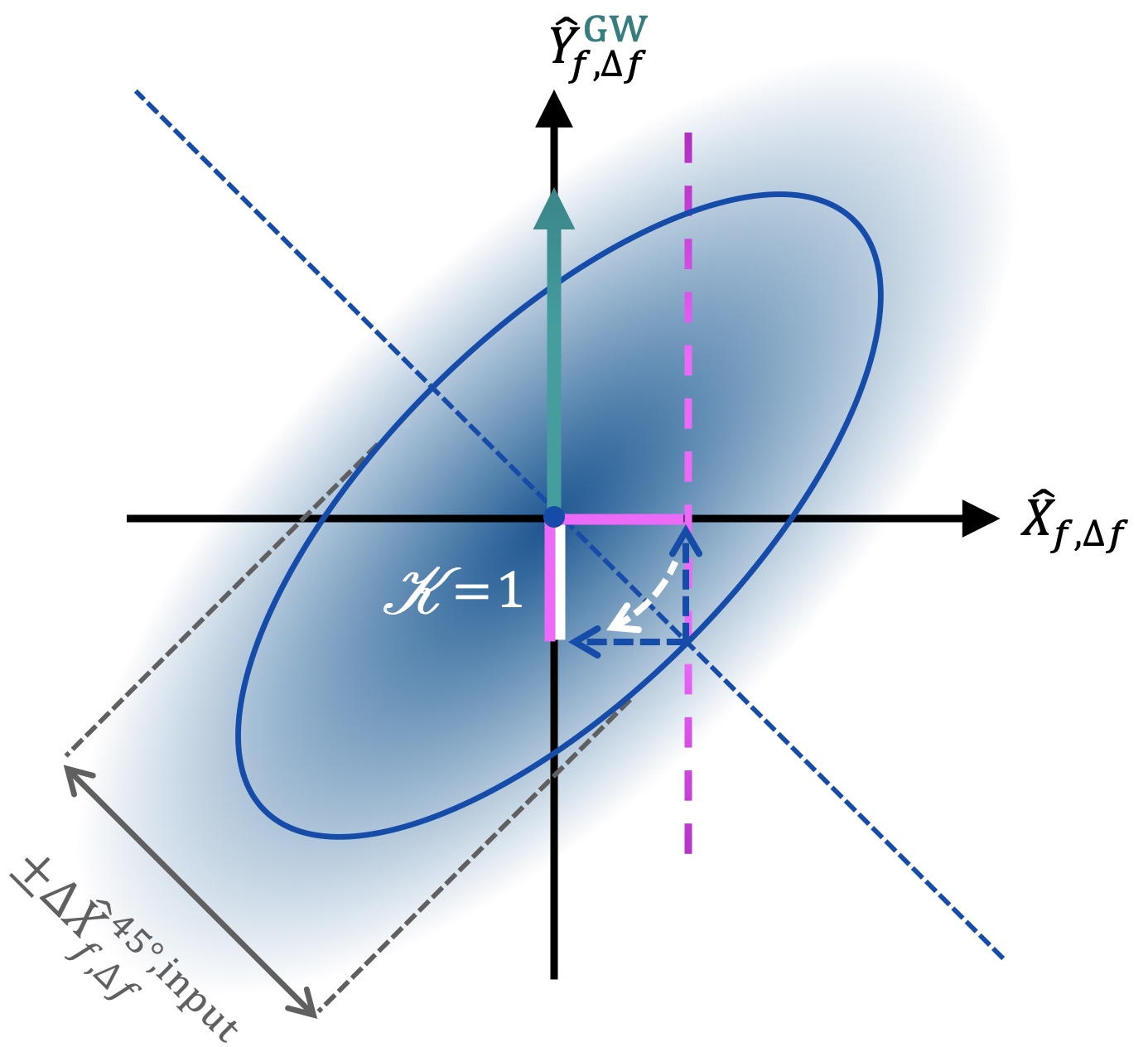}
  \vspace{-1mm}
  \caption{Remaining uncertainty in $\hat Y$ --  This figure completes the illustration in Fig.\,\ref{fig:5} by its remaining uncertainty in $\hat Y$.
  Shown here is that the same radiation pressure as in the figure above has also a subordinate anti-correlated component in the uncertainty of $\hat Y$. The radiation pressure back action produces another anti-correlated component in the uncertainty of $\hat Y$ of the same magnitude (curved dashed arrow). The two anti-correlated components constructively interfere to a magnitude that is $\sqrt{2}$ larger than the squeezed standard deviation. Since at $f_{\rm SQL}$ the standard deviation of the quantum uncertainty in $\hat Y$ is $\sqrt{2}$ larger than the ground state standard deviation, the full squeeze factor is retained. Note that at higher (lower)  frequencies the injected squeeze angle needs to be lower (higher) to retain the full squeeze factor \cite{Jaekel1990}. In the recent experiment at LIGO, the injected squeeze angle was $24^\circ$, $35^\circ$, and $46^\circ$ for which the inferred quantum noise in the observation of gravitational waves surpassed the SQL, see Figs.\,2 and 3 in Ref.\cite{Yu2020}.
}
  \label{fig:6}
\end{figure}

\section{Coupling mechanical and optical uncertainties} \label{sec:2}
Macroscopic mechanical oscillators promise to enable tests of quantum physics on a macroscopic scale. 
Furthermore, solid bodies couple to the gravitational force, which enables tests of quantum gravity theories in the weak field regime.
The idealised model of such a mechanical system is a movable rigid mirror that is suspended as a pendulum and probed by a monochromatic laser beam.
The light propagates over some distance, reflects off the mirror, and propagates further where it is photoelectrically detected.
The total optical path length depends on the precise position of the mirror surface. 
The phase of the light when detected thus carries information about even tiniest position changes of the mirror. 
A laser interferometer measures this phase change in comparison to the optical length of a reference path, see Fig.~\ref{fig:2}.
There are two quantum systems involved, both defined by their own uncertainty: the continuously measured optical modes and the mechanical modes of the mirror that interacted with the optical modes. 
Usually, the optical modes are in rather pure states, i.e.~in ground state or in displaced ground states when a GW signal is observed or disturbances present. 
The mechanical modes are in thermal states because they are in thermal equilibrium with the environment. 
The high Q-factor of the suspension leads to a characteristic time during which the {thermal decoherence} (thermalisation) takes place. 
The higher the Q-factor, the slower the {decoherence}.

An optical field with fluctuations over a broad spectrum of frequencies is traditionally described in terms of spectral Fourier components of amplitude ($X$) and phase quadrature fluctuations ($Y$) at frequency $f$: $\hat{a}_{X,Y}(f)$ respectively \cite{Caves1985a}.
The actual observables $\hat X_{f,\Delta f} (t,\Delta t)$ and $\hat Y_{f,\Delta f} (t,\Delta t)$, however, are the Fourier transformed averages over the resolution bandwidth $\pm \Delta f$ of these quantities.
When the mode is in a pure quantum state, the uncertainties of its phase and amplitude quadratures obey Heisenberg uncertainty relation: $\Delta\hat X_{f,\Delta f} (t,\Delta t) \cdot \Delta\hat Y_{f,\Delta f} (t,\Delta t) \ge 1/4$, where the actual number on the right depends on the normalisation. 
For a coherent state, the uncertainties are equal, i.e.~$\Delta\hat X_{f,\Delta f} (t,\Delta t) = \Delta\hat Y_{f,\Delta f} (t,\Delta t) = 1/2$. 
It is possible to create a state, in which one of the uncertainties is decreased at the expense of the other, i.e.~$\Delta\hat X_{f,\Delta f} (t,\Delta t) = e^{-r}\!/2$ and $\Delta\hat Y_{f,\Delta f} (t,\Delta t) = e^{r}\!/2$, while also obeys the uncertainty relation.
Such state is called `squeezed' \cite{Walls1983}, with $r$ the squeeze factor \cite{Stoler1970}. 
In general, a state can also be squeezed for any linear combination of the amplitude and phase quadratures. 
The `squeeze angle' \cite{Stoler1970} is usually defined with respect to the amplitude quadrature.

Whenever light is reflected off a mirror, it exerts radiation pressure on it.
The corresponding force accelerates the mirror, i.e.~transfers momentum.
This radiation-pressure force depends on the amplitude of the input light field $\hat X^{in}_{f,\Delta f} (t, \Delta t)$. 
Since this amplitude has some uncertainty, it is transferred onto the mirror's momentum $\hat p^{in}_{f,\Delta f} (t, \Delta t)$, which influences at later times its position $\hat x^{in}_{f,\Delta f} (t, \Delta t)$.
This phenomenon is called quantum back-action (QBA) when sensing the mechanical motion with light.
When continuously sensing (monitoring) the mirror position, QBA increases the position uncertainty, which results in additional quantum noise in the measurement record. 
This quantum radiation-pressure noise (QRPN)\,\cite{Caves1981} may hinder the sensitivity of various detectors operating in quantum domain, such as gravitational-wave detectors.
QRPN was first observed in microscopic systems \cite{Murch2008,Purdy2013}, and only recently in GW observatories, see the next section.

In the following, we describe quantum radiation pressure formally in a quasi-stationary case.
The Fourier components of the quantum radiation pressure force are given by the amplitude quadrature of the incident light $\hat{F}_{\rm rp}(f) = \hbar \alpha \, \hat{a}_X(t) \propto \sqrt{P_0} \,\hat a_X(f)$, where $\alpha$ is the normalized average amplitude, which depends on the average optical power $P_0$ of the monochromatic carrier light at the mirror.
The force leads to a mechanical displacement $\hat{x}_{\rm rp}(f) = \chi(f)\hat{F}_{\rm rp}(f)$, where $\chi(f)$ is the complex valued mechanical response function.
The phase quadrature of the light field reflected of a movable mirror $\hat b_Y(f)$ picks up the displacement of the mechanical oscillator, which in turn is driven by the radiation-pressure force.
Fluctuations of the quadratures in the linear approximation, i.e.~when the displacement is small relative to the wavelength, and the fluctuations are small relative to average field amplitude, can be expressed by two equations (ignoring irrelevant phase factors): 
\begin{align}
  \hat{b}_X(f) =& \; \hat{a}_X(f);\\
  \hat{b}_Y(f) =& \;\hat{a}_Y(f) - \alpha \left(\hat{x}_{\rm rp}(f) + \hat{x}_{\rm sig}(f)\right) \nonumber \\
  = & \; \hat{a}_Y(f) - \mathcal{K}(f) \hat{a}_X(f) - \alpha\chi(f)F_{\rm sig}(f) \, , 
\label{eq:bY}
\end{align}
where $F_{\rm sig}(f)$ is the signal force and {$\mathcal{K}(f) = \hbar \alpha^2 \chi(f) $ is the optomechanical coupling factor (also called Kimble factor~\cite{Kimble2001}).
In a more general case, $\mathcal{K}(f)$ also includes the effects of the optical cavities in the arms (see for details e.g.~\cite{Danilishin2012}).

Eq.\,(\ref{eq:bY}) describes the emergence of quantum back action, which is illustrated in Fig.\,\ref{fig:3}.
The first term corresponds to the measurement (shot) noise (QMN), and the second term -- to quantum back action noise (QBN).
An important property of quantum back action can be seen in this equation: radiation-pressure force quantum-correlates phase quadrature of the output field with amplitude quadrature of the incoming field.
The output state becomes \textit{ponderomotively} squeezed~\cite{Braginsky1967,Kimble2001}.
This quantum correlation can be seen from the fact that for some linear combinations of $\hat{b}_X(f)$ and $\hat{b}_Y(f)$ the radiation pressure back action cancels, see the illustration in Fig.\,\ref{fig:4}.
Unfortunately, a GW observatory with just a single photo diode cannot benefit from ponderomotive squeezing alone~\cite{Kimble2001}.
In contrast, the ponderomotive anti-squeezing results in the quantum radiation pressure noise.\\

\emph{The standard quantum limit -- }
The total quantum noise of a measurement device is given by the sum of quantum back-action noise and quantum measurement noise, where both are normalised to the signal strength \cite{Braginsky1995}. 
With this normalisation, doubling the light power doubles the signal-normalised QBN and halves the signal-normalised QMN. 
Without correlations of QBN and QMN, their sum is minimal if they are balanced. 
For every value of the light power, there is exactly one frequency, at which the sum of QBN and QMN is minimal. 
The spectrum of all minima defines the standard quantum limit (SQL) of a measurement device.
It represents the lowest quantum noise for continuous measurements without exploiting quantum correlations.

Continuous measurements allow for a spectral analysis of the signal. 
The sensitivity of a corresponding measurement device can be quantified by its `signal-equivalent noise spectral density'. 
The signal-equivalent QBN spectral density decreases with frequency above the mechanical resonance. 
For this reason, also the SQL decreases with frequency, and a quantum noise limited GW observatory that does not exploit quantum correlations only reaches the SQL at a single frequency. 
This frequency changes, however, when the optical power {is changed}. 
The total signal-equivalent quantum noise spectral density (normalized to displacement) is given by
\begin{align}
  &S_x(f) = \frac{x^2_{SQL}}{2} \left(\frac{1}{\mathcal{K}(f)} + \mathcal{K}(f) \right);\label{eq:sens} \\  
  &x_{\rm SQL}(f) = \frac{1}{2\pi}\sqrt{\frac{8\hbar}{Mf^2}},
\end{align}
where $x_{\rm SQL}(f)$ is the SQL for the free mirror with a mass of $M$.
The SQL is achieved at the frequency, where $\mathcal{K}(f_{\rm SQL})=1$.
Eq.\,(\ref{eq:sens}) also holds if squeezed light is injected for squeezing the output light's amplitude quadrature spectrum. 
In this case $\mathcal{K}(f)$ needs to be multiplied by $e^{2r}$, and the SQL cannot be overcome.
The SQL can be overcome, however, by employing quantum correlations or quantum non-demolition measurements~\cite{Jaekel1990,Kimble2001,Danilishin2012}.
The simplest approach is the injection of a squeezed light with a frequency independent squeeze angle $\theta \neq 0^\circ$.
In this case, Eq.\,(\ref{eq:sens}) is not valid.
This was recently demonstrated in LIGO, as we summarise in the next section.

\begin{figure}
  \includegraphics[width=0.9\columnwidth]{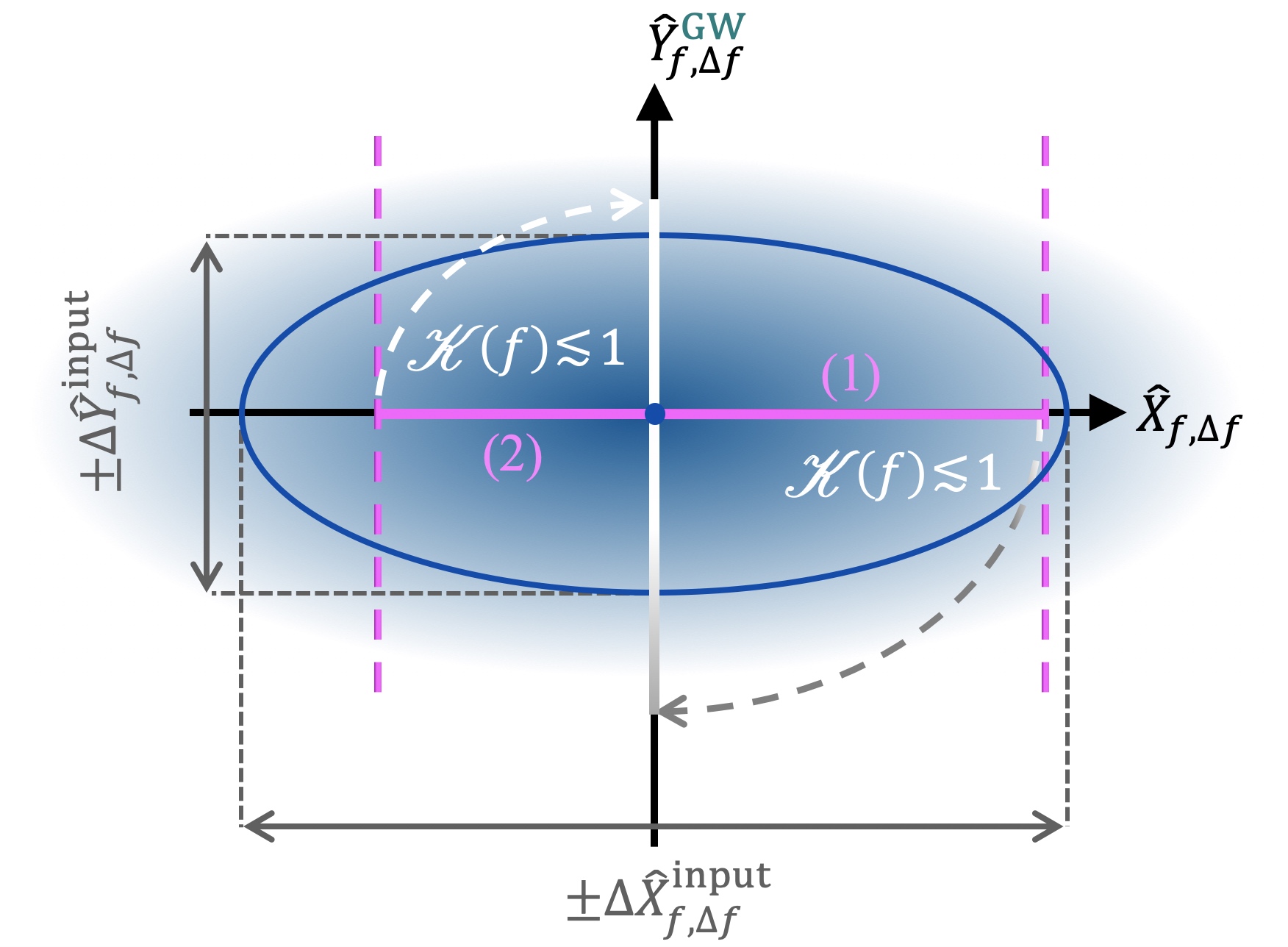}
  \vspace{-1mm}
  \caption{Quantum back action due to the input light's anti-squeezing --  The fussy elliptical area represents the quantum uncertainty of the squeezed vacuum state with a squeeze angle of $\theta = 0^\circ$. The anti-squeezed uncertainty of the light produce a larger back action than the ground state in Fig.\,\ref{fig:3}. This is possible even if the optomechanical coupling factor $\mathcal{K}$ is smaller than that at the SQL, as shown here. 
}
  \label{fig:7}
\end{figure}

\begin{figure*} 
\floatbox[{\capbeside\thisfloatsetup{capbesideposition={left,top},capbesidewidth=6.6cm}}]{figure}[\FBwidth]
{\caption{Observation of quantum back-action in Virgo -- Shown are three measured spectra (solid lines) and corresponding quantum noise models \cite{Acernese2020}. The trace levelled in the centre (black) was the observatory noise without squeezed vacuum injection. When squeezing was injected with a squeeze angle of $\theta = 0^\circ$, the shot noise on the photo diode was clearly squeezed above 200\,Hz (red trace). For this squeeze angle, the differential quantum radiation pressure noise (QRPN) in the arm resonators increased, and QRPN became visible between 30\,Hz and 40\,Hz. 
With $\theta = 90^\circ$, the shot noise was anti-squeezed (blue trace), and the increases shot noise elevated the total noise between 30\,Hz and 40\,Hz. The grey trace represents the total non-quantum noise. Reproduced with permission from Phys. Rev. Lett. 125, 131101 (2020) under CC BY \cite{Acernese2020}, with additions to the original.
}}
{\includegraphics[width=10.7cm]{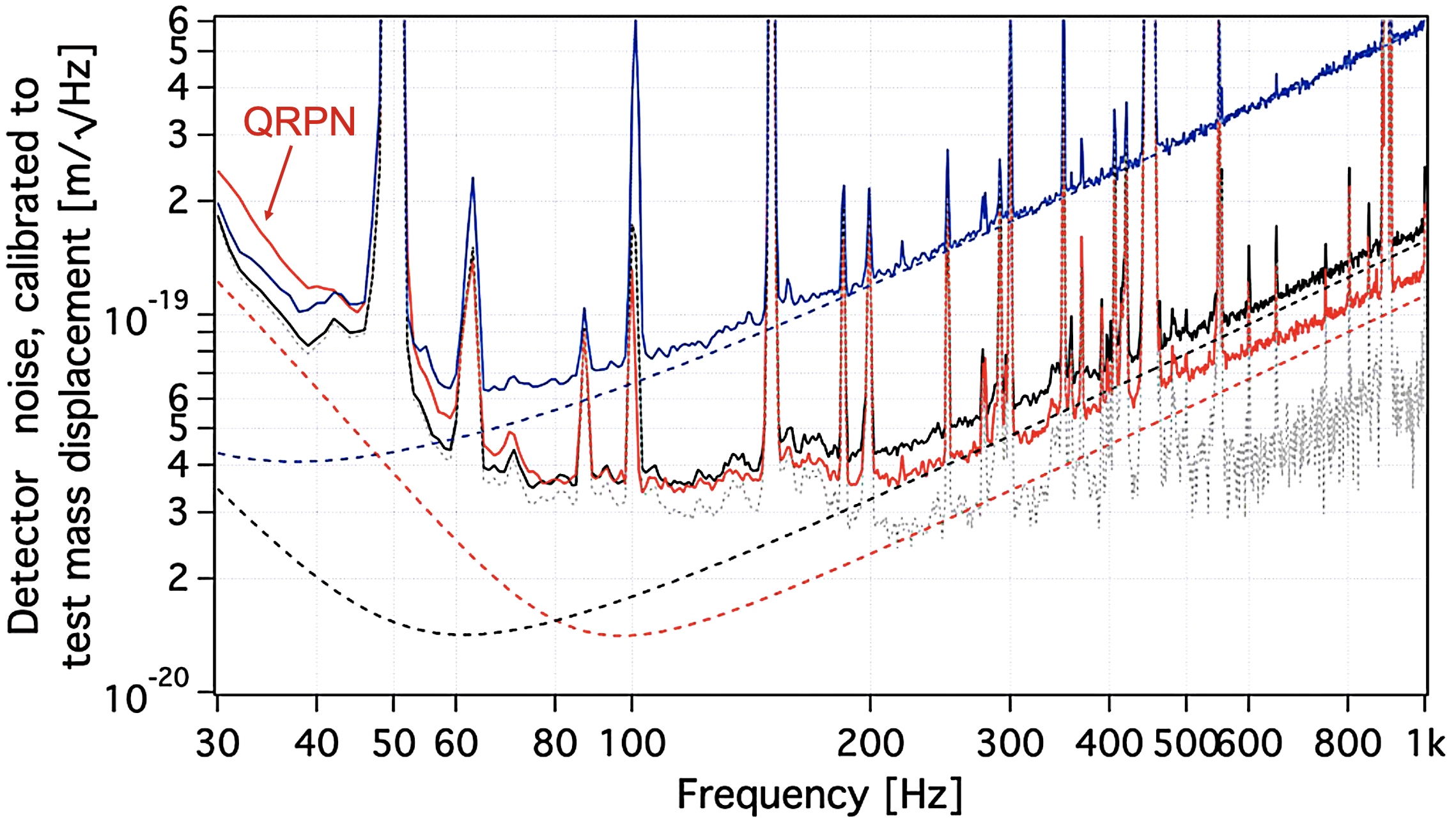}  \label{fig:8}}
\vspace{2mm}
\end{figure*}

\section{Observation of optomechanical coupling in GW observatories} \label{sec:3}
Recent experiments with LIGO and Virgo revealed coupling of the quantum uncertainties of the light with the differential motion of the four 40\,kg-sized mirrors of the two arm resonators. 
In Virgo, the squeezed vacuum state was injected with a squeeze angle $\theta = 0^\circ$ as it was done previously in GEO\,600\cite{LSC2011}, LIGO\cite{Tse2019}, and Virgo\cite{Acernese2019}. 
With a slightly increased laser power, the noise of the output light showed a significant contribution of quantum back action at frequencies between 30\,Hz and 40\,Hz. 
The back-action in terms of quantum radiation pressure noise was due to the anti-squeezed quadrature of the injected squeezed vacuum states. 
The illustration of the setting is given in Fig.\,\ref{fig:7}. 
It is similar to Fig.\,\ref{fig:3}, but due to the injected squeezed vacuum states the quantum uncertainty in the radiation pressure is anti-squeezed and visible even in the presence of non-quantum noise sources.
Fig.\,\ref{fig:8} shows measurements of the square root of the Virgo noise spectral density calibrated to differential arm length in ${\rm m}/\!\sqrt{{\rm Hz}}$. 
The effect of quantum radiation pressure noise (QRPN) was directly observed through the elevation of the upper trace between 30\,Hz and 40\,Hz with respect to the other traces. 
Notably, the QRPN effect was not masked by non-quantum noise sources. (The same effect can also be seen to a small extent in Fig.\,1 of Ref.\cite{Tse2019} and in Fig.\,2 of Ref.\cite{Acernese2019}.)

The spectral density of the squeezed noise generated in Virgo was 13.8\,dB below that of the ground state noise (vacuum noise). 
The total quantum efficiency was about 54\%, i.e.~46\% of the energy in the squeezed vacuum states were lost and the pure squeezed states got mixed with the corresponding contribution of the ground state. 
The loss reduced the anti-squeezing from 13.8\,dB (a factor of about 24 above the variance of the ground state) to 11.3\,dB ($\sim$13.4). 
The square root of this factor ($\sim$3.7) corresponds to the factor between the black and the blue trace at the high frequencies in Fig.\,\ref{fig:8}. 
The same loss reduced the squeezing from $-13.8$\,dB to about $-3$\,dB ($\sim$0.5), and a phase squeezed trace (red) a factor of about 0.7 below the vacuum noise (black) at the high frequencies in Fig.\,\ref{fig:8}. 
In reality both values were closer to unity because of the underlying non-quantum noise (grey).
The observed QRPN was thus produced by a mixed state with a product of the uncertainty standard deviations that was a factor of $0.7 \times 3.7 = 2.6$ above minimum uncertainty. 
The lower the factor is, the more `pure' is the optomechanical coupling and the `more quantum' is the observed QRPN.

A more solid criterion for `quantumness' of optomechanical coupling is given by the standard quantum limit. 
If the measured total spectral density shows with statistical significance (over some finite frequency band $f \pm \Delta f$) that the overall quantum noise is below the SQL, optomechanical quantum correlations (OMQC), i.e.~quantum correlations between the uncertainties of the light field and the mirror motion are proven.
Therefore, the SQL serves as a useful benchmark for quantifying the possibility to measure quantum-mechanical effects for the specific optomechanical Fourier mode with eigenfrequency $f \pm \Delta f$.
This mode could then be used to study the foundations of quantum theory or gravity, as we discuss in the next section.

With the input squeeze angle set to $\theta = 0^\circ$, today's GW observatories cannot observe a quantum noise below the SQL at any frequency. 
The reason is that surpassing the SQL requires the actual exploitation of quantum correlations, but this is impossible with just a single photo diode in the output port and a zero squeeze angle in the input. 
The quantum noise can surpass the SQL over a broad frequency range if the input squeeze angle is set to an optimal (non-zero) value and the output optics supplemented by filter cavities, which provide a frequency dependent quadrature rotation, and a balanced homodyne detector, which is able to detect an arbitrary fixed quadrature angle \cite{Kimble2001}.
The simplest approach to surpass the SQL is the optimisation of the input squeeze angle without any further changes to the optics at the output port of a today's GW observatory. 
This approach allows for suppressing the quantum noise below the SQL over a finite but not too narrow frequency band around a well-defined frequency ($f_{\rm OMQC}$). 
This experiment was done in LIGO~\cite{Yu2020}. 
The total spectral density was considerably above the SQL, but its reduction over the characteristic frequency band when optimising the squeeze angle allowed the inference of the overall quantum noise below the SQL.
The results are shown in Figs.\,\ref{fig:9} and \ref{fig:10}. 
The quantum noise was demonstrated to be below the SQL for three different squeeze angles.
\begin{figure*} 
\floatbox[{\capbeside\thisfloatsetup{capbesideposition={left,top},capbesidewidth=4.0cm}}]{figure}[\FBwidth]
{\caption{Experimental inference of mirror/light quantum correlation in LIGO -- Squeezed vacuum states with a squeeze angle of $\theta = 35^\circ$ were injected. The noise spectral density around 40\,Hz was below the one without squeezing injection. A total quantum noise below the SQL was inferred by subtracting non-quantum noise, which was determined by reference measurements. The smooth solid lines represent quantum noise models. Reproduced with permission from H. Yu et al., Nature 583, 43 - 47 (2020). Copyright 2020 Springer Nature. 
}}
{\includegraphics[width=13.3cm]{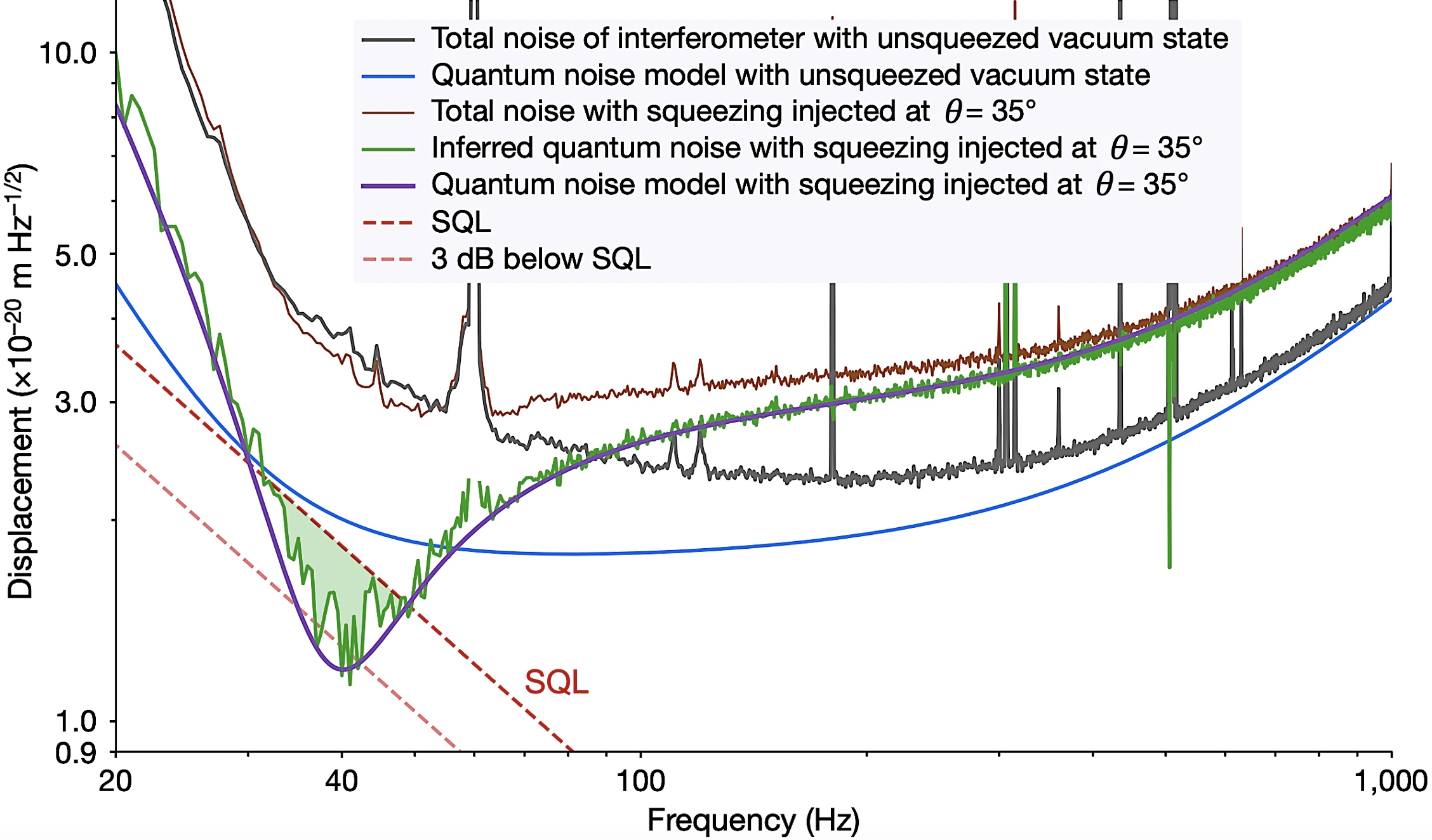}  \label{fig:9}}
\vspace{2mm}
\end{figure*}

\begin{figure*}
  \includegraphics[width=18cm]{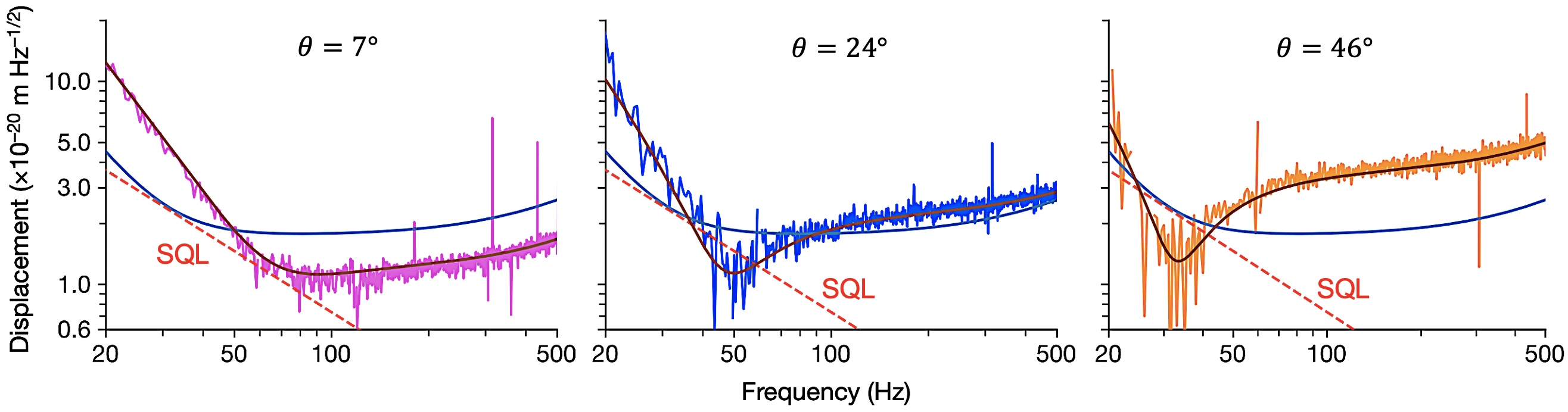}
  \caption{Inferred mirror/light quantum correlation for different squeeze angles -- Squeezed vacuum states with squeeze angles of $\theta = 7^\circ,\, 24^\circ, \,46^\circ$ were injected. For the two larger angles, a total quantum noise below the SQL was inferred by subtracting non-quantum noise, which was determined by reference measurements. The inference for $\theta = 46^\circ$ at a frequency of $f_{\rm SQL,\,LIGO} \approx 30$\,Hz almost resembles the illustration in Figs.\,\ref{fig:5} and \ref{fig:6}. Reproduced with permission from H. Yu et al., Nature 583, 43 - 47 (2020). Copyright 2020 Springer Nature. 
  }
  \label{fig:10}
\end{figure*}

The overall displacement-normalised quantum noise spectral density for injected squeezed light with squeeze angle $\theta$ can be calculated by the following expression
\begin{align}
  S_x(f) = & \frac{x^2_{SQL}}{2} \left(\frac{1}{\mathcal{K}(f)} + \mathcal{K}(f) \right)\times \nonumber \\
  &\times \left[e^{-2r} \cos^2 \left(\theta - \vartheta(f) \right) + e^{2r} \sin^2 \left(\theta - \vartheta(f) \right)\right];\\
  \mathcal{K}(f) = &\frac{2 \mathcal{J}\gamma}{(2\pi)^3 f^2 \left(\gamma^2 + f^2\right)}; \quad \mathcal{J} = \frac{8\pi\nu_0P_{\rm arm}}{M L c};\\
  \vartheta(f) = &\arctan \mathcal{K}(f),
\end{align}
where $\gamma$ is the detector bandwidth, $L$ is the arm length, $P_{\rm arm}$ is the optical power in the arm cavities, $\nu_0$ is the frequency of laser, $c$ is the speed of light, $M$ is the  mass of an end mirror.
For example, at the frequency, where the noise touches the SQL, $\mathcal{K}(f_{\rm SQL}) = 1, \vartheta(f_{\rm SQL})=45^\circ$. 
Then, choosing the injected squeeze angle of $\theta = \vartheta(f_{\rm SQL})$ allows to dip below the SQL exactly by the amount of available squeezing:
\begin{equation}
  S_x(f_{\rm SQL}) = x^2_{SQL}e^{-2r}.
\end{equation}
This can be seen in Fig.\,\ref{fig:10}, right.

A direct observation of the total noise spectral density below the SQL was not possible in LIGO because there are multiple sources of technical noise, which contribute at a level above the SQL.
Most notable sources of noise are: thermal motion of suspensions~\cite{Cumming2012} and surfaces~\cite{granata2020progress} of test masses, seismic vibrations coupling to the motion of the mirror~\cite{vanHeijningen2019} as well as control noises. 

These observations of quantum effects in the detectors are just the first steps towards application of advanced quantum technology for gravitational-wave observatories, which are reviewed in detail in Refs.~\cite{Danilishin2012,Danilishin2019}.
Nonetheless, even this progress opens the path towards the tests of fundamental quantum mechanics and gravity, as we discuss in the next section.

\section{Quantum correlations and tests of spontaneous decoherence} \label{sec:4}
\subsection{Non-Gaussian optomechanical states for probing decoherence}
Macroscopic systems are generally not observed in pure quantum states, such as superposition states.
The question of why the world appears ``classical'' to us, is understood as the consequence of decoherence mechanisms, which quickly
destroy macroscopic superpositions~\cite{Zurek2003}.
These mechanisms usually considered obey the rules of standard quantum mechanics. 
Therefore, they neither allow to resolve the measurement problem of quantum mechanics, nor shine light on the connection between quantum mechanics and gravity.
Spontaneous decoherence models attempt to resolve either problem, or both, by modifying quantum mechanics and introducing additional terms in the Schr\"odinger equation.
This manifests itself as a source of spontaneous decoherence, which for large objects is stronger than other typical sources of environmental decoherence.
Two most prominent examples of these spontaneous decoherence models are continuous spontaneous localisation (CSL)~\cite{Ghirardi1990, bassi2003dynamical, Bassi2013} and Di\'{o}si-Penrose (DP) models~\cite{Diosi1987, Diosi1989, Penrose1996, Penrose1998}.
In the CSL models, the decoherence occurs due to an additional stochastic force, which acts on all objects, with its strength being stronger for larger objects.
In the DP model, the decoherence is caused by the gravitational self-interaction of different parts of a wave-function.
Both mechanisms, although different in nature, are related to the mass of objects.
The potential relevance of gravity in the emergence of a `classical' everyday world was pointed out by Frigyes Karolyhazy already in the 1960s~\cite{Karolyhazy1966}, and a recent review is given in \cite{Bassi2017}.

In 2003, Marshall, Simon, Penrose, and Bouwmeester proposed an experiment~\cite{Marshall2003} that was intended to realise a mechanical system having a position uncertainty with two probability maxima that were separated by the width of the ground state uncertainty.
The envisioned mechanical system was a mirror with a volume of $10\,{\rm \mu m}$ cubed and a mass of $m = $5\,ng that constituted one end of a 5\,cm long cavity in one arm of a Michelson interferometer, see Fig.\,\ref{fig:10}. The second cavity mirror was much heavier and not movable. The mesoscopic mirror was mechanically suspended with a resonance frequency of $f_m = 500$\,Hz, which resulted in a ground state half-width position uncertainty of $\Delta x_m =  \sqrt{\hbar / (4 \pi m f_m)} \approx 6 \cdot 10^{-13}$\,m. The excitation of the cavity mode to a single photon Fock state $|1\rangle$ was calculated to be sufficient to produce a radiation pressure induced displacement of the tiny mirror by about the same size. The second arm of the interferometer contained another cavity of identical optical parameters but with two immovable macroscopic mirrors.\\ 
Coupling a single photon to a decoherence-free interferometer entangles the quantum uncertainty of the mirror position with the optical fields in the two arm cavities. 
If the position of the mirror shows some spontaneous decoherence, the mirror either experiences the full radiation pressure, is displaced, and the photon is in cavity A, or the mirror does not experiences any radiation pressure, is not displaced, and the photon is in cavity B. Ensemble measurements without decoherence would result in 
self-interference of the photon into one of the output ports. Ensemble measurements with decoherence of the mirror position would always result in a photon that is localised to one of the arm cavities. Interference would not possible any more. An ensemble measurement would find the out-coupled single photons always randomly distributed in the output ports, regardless what the precise differential arm length of the interferometer was. Note, that the entire ensemble measurement is \emph{conditioned} on a single `click' of either detector D1 or D2.

\begin{figure}
\vspace{1mm}
  \includegraphics[width=6.5cm]{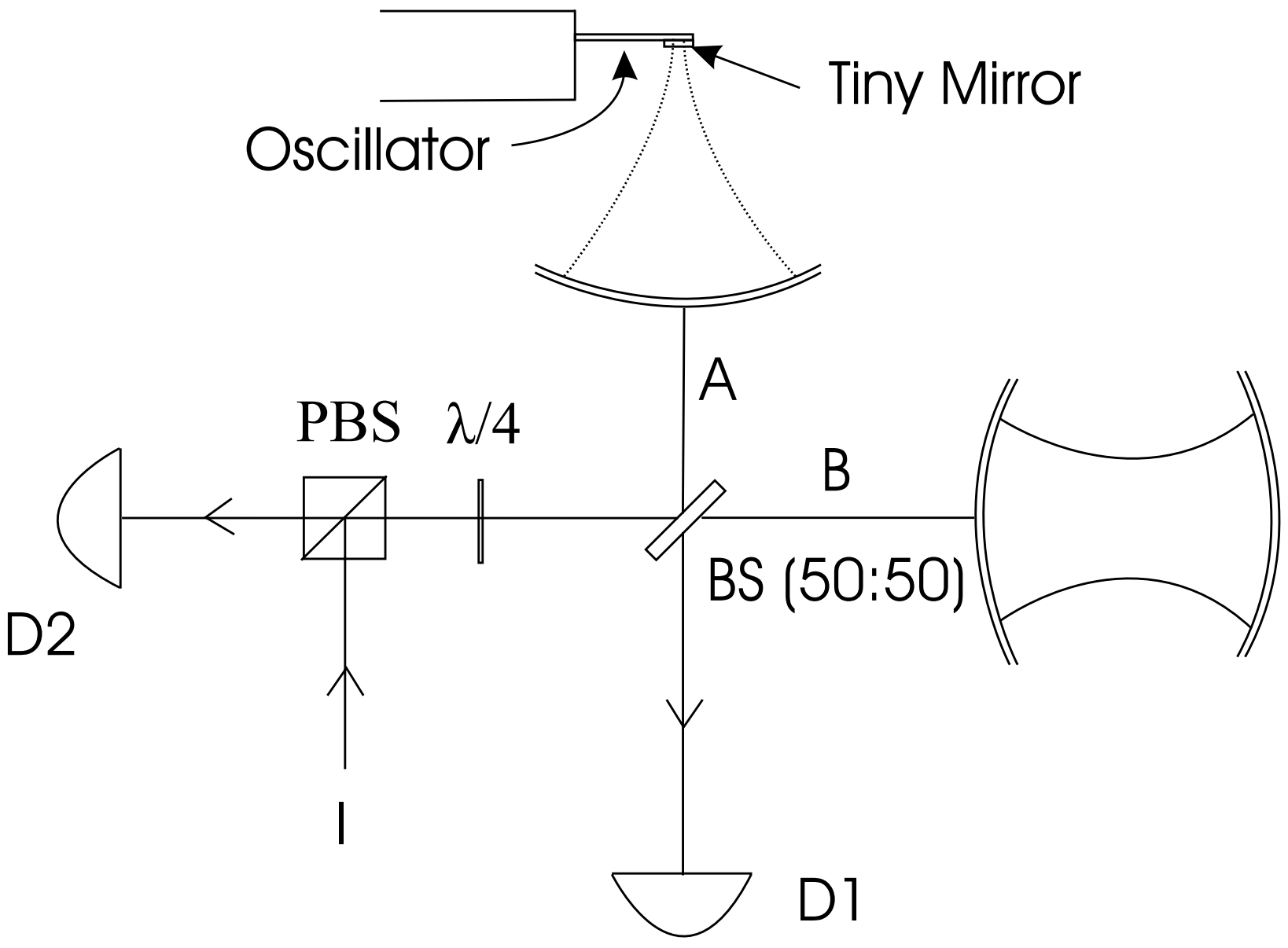}
  \caption{Setup proposed in Ref.\cite{Marshall2003} --  An optical single-photon state $|1\rangle$ enters a Michelson interferometer with high-finesse arm cavities. The cavity in arm A has a tiny end mirror mounted on a micromechanical oscillator, and the radiation pressure places the mirror in a non-Gaussian state. If decoherence localizes the mirror on time scales shorter than the cavity mode life time, it either experiences full or zero radiation pressure. This means, it localises the photon in one or the other arm. Interference effects vanish due to the localisation. D1, D2: single-photon detectors. Reproduced with permission from W. Marshall, C. Simon, R. Penrose, and D. Bouwmeester, Phys. Rev. Lett. 91, 130401 (2003). Copyright 2003 by the American Physical Society.
  }
  \label{fig:11}
\end{figure}

The non-Gaussian position uncertainty of a mesoscopic mirror proposed in~\cite{Marshall2003} has two local maxima with a separation of the order of the motional ground state. The dimension of the mirror, however, is many orders of magnitudes larger. 
Preparing mesoscopic objects in superpositions of two positions that are separated by more than the objects diameter is far beyond current technology. (Such a state could rightly be called a Schr{\"o}dinger cat state.)
Theoretical work showed that reasonable assumptions on gravitational decoherence mechanisms might involve rather large decoherence time scales. 
Experiments with state of the art technology will be rather limited by \emph{environmental} decoherence mechanisms and thus not sensitive to most of the gravitational decoherence mechanisms discussed so far~\cite{Bose1999, Bernad2006, Adler2007, Kleckner2008a}.
Due to this, recent works focused on more microscopic systems, involving matter-wave interferometry~\cite{Bose2017,Nimmrichter2011} or Bose-Einstein condensates~\cite{Howl2019}.

Ensemble of non-Gaussian states of light have been usually produced from spontaneously produced photon pairs. 
Conditioned on the successful detection of one of the photons, a Fock-1-states can be created and phase space tomography performed \cite{Lvovsky2001}. 
Recently, also phononic Fock-1-states were produced \cite{Hong2017,Riedinger2018,Marinkovic2018}.
A theoretical analysis suggested to prepare the mirror motion in GW observatories in non-Gaussian states by injecting optical Fock-1-states into the signal output port \cite{Khalili2010} following previous theoretical work \cite{Zhang2003b}.
The quantum uncertainty of the single photon is optically amplified by the intense light in the arms and produces a significant non-Gaussian quantum radiation pressure force on the mirrors and thus a non-Gaussian quantum state of the joint mirror motion.
It was pointed out that some experiments on testing stochastic gravitational wave-function collapse large masses are not always preferable, despite stronger coupling to gravity~\cite{Nimmrichter2014}.
However, bringing massive mechanical oscillators into quantum regime allows to test the limits of quantum theory as well as gravity.

\subsection{Gaussian optomechanical states for probing decoherence}
Both, nonclassical Gaussian and non-Gaussian states are viable probes for testing spontaneous decoherence. 
On both nonclassical states, any decoherence is noticeable through a redistribution of the uncertainty.

Gaussian squeezed states were recently used as probe systems to quantify an environmental decoherence process \cite{Vahlbruch2016}. The decoherence of interest was the escape of photons in the course of their detection by photodiodes. The measurement utilised the fact that photon escape deteriorates the measured state's purity.
 The product of the standard deviations of the squeezed and anti-squeezed uncertainties increased successively. By the direct observation of a 15\,dB squeezed state and independent reference measurements of other loss sources, it was possible to quantify the photon escape during measurement to about 0.5\% with an error bar of the same size. The photo diodes' quantum efficiency was thus measured to $(99.5 \pm 0.5)$\%.

One could argue that a non-Gaussian state is more sensitive to decoherence and is therefore better suited as a probe system. 
However, in this case the same state is also more sensitive to decoherence due to disturbances from the environment, which offsets such an advantage. Nevertheless, non-Gaussian states might be more sensitive to some models of gravitational decoherence, owing to distinct spatial maxima in a wavefunction. Gaussian states, on the other hand, offer a possibility to continuously monitor the dynamics of a quantum state, thus observing the evolution of multiple modes with time under the action of various of sources of decoherence.

The generation of Gaussian quantum-correlated optomechanical as well as purely mechanical states was investigated theoretically by several authors. 
To the best of our knowledge, the first theoretical consideration of an optomechanical quantum correlated state was done by Bose, Jacobs and Knight \cite{Bose1997,Bose1999}. 
They considered a macroscopic mirror in a cavity and found that it may be placed in a stationary Schr\"odinger-cat-like state by a quadrature measurement of the light field after interaction. 
They also investigated whether such a state could be used to detect gravitationally induced decoherence, and concluded that this was way beyond technology at the time. 
Subsequent research found that stationary entanglement between vibrational modes of two cavity mirrors, with an effective mass of the order of micrograms, can be generated by means of radiation pressure \cite{Vitali2007}.
Furthermore, it was proposed that also stationary entanglement can be generated between the motion of the four mirrors in a LIGO-type GW observatory \cite{Mueller-Ebhardt2008,Mueller-Ebhardt2009,Schnabel2015}. 
Here, the generated entanglement is with respect to position and momentum of the entire pendulum suspended mirrors, however, not for the resonance mode of the pendula but for overtone modes of the frequency band $f \pm \Delta f$, for which the sum of all classical noise contributions was below the spectral density of the SQL. 
The continuous monitoring of the mechanical mode allowed for the measurement of the thermally driven random walk and for the conditioning (referencing) of the quantum uncertainties with respect to this random walk \cite{Mueller-Ebhardt2009diss}. 

Using an optomechanical system in a nonclassical Gaussian states for probing decoherence processes is done via continuous monitoring of the states uncertainty under the continuous influences of optomechanical coupling and decoherence. Crucial is the reconstruction of the thermally driven random walk and subtracting it from the measurement data. The quantum uncertainty is then revealed \emph{conditioned} on the knowledge of this classical trajectory.
Unlike the conditioning on the click of a photon detector in non-Gaussian state preparation, here a full measurement record is taken into account to estimate the current state.
The classical random walk requires optimal processing of the information with the use of Wiener (or Kalman) filtering, which includes the complete model of the system.
This allows to trace the random evolution of the state, as well as its uncertainty, as it was theoretically proposed~\cite{Mueller-Ebhardt2008}, further theoretically investigated~\cite{Mueller-Ebhardt2009,Hoff2016,Hofer2015,Wieczorek2015,Murch2013,Gut2020} and recently demonstrated experimentally~\cite{Rossi2019,Magrini2021}. 

Continuous monitoring and conditioning on the knowledge of the mean random walk provide the key feature that allows to check for the changes in the evolution of the quantum state, and not only for stationary effects.
This can allow to directly probe for modifications to the Schr\"odinger equation, which governs the evolution of the system, and potentially see the dynamical effects of decoherence, possible non-linearities in the equation or signatures of classical gravity~\cite{Helou2017}.
Usually, in order to observe these changes, the system should be allowed to evolve freely, without interaction with the probe light.

The state after the free evolution is then verified and compared to the predicted dynamics, as was demonstrated experimentally in~\cite{Rossi2019}.
At this stage any deviation from the predicted dynamics becomes visible.
The verification stage is required, since the state obtained during the preparation stage depends on the validity of the model for the system.
The main difficulty of this process is in the need for sub-SQL sensitivity for revealing non-classical features~\cite{Mueller-Ebhardt2008}.
The verification step is crucial, but also the most difficult, since it requires a back-action evasion protocol in order to reveal non-classical features in the state~\cite{Miao2010,Danilishin2013}.

Continuous monitoring of a nonclassical Gaussian optomechanical quantum state can allow for detecting decoherence
effects, which might be not visible with the use of non-Gaussian states.
For example, if the entanglement is prepared for the mirrors of LIGO for the 50\,Hz Fourier mode,  and then let to evolve freely, it would decohere due to thermalization over the timescale of $\approx 3ms$.
One of the gravitational decoherence models suggest the characteristic timescale of $\tau_{gd}\approx1\mu s$ (based on the consideration in \cite{Miao2010}). 
Therefore, if we verify the state after $\sim 100 \mu s$, and find no entanglement between the test masses, we can be confident that this decoherence occurred due to the unknown mechanism, possibly gravitational.
The timescales of the experiment are easily adjustable, since it is done in continuous regime.
Similar experiment with non-Gaussian states would be difficult, since the preparation of the state is single-shot.

Although the dynamics of the entangled state would allow for testing for a wide parameter range of decoherence models, conditional state preparation is experimentally challenging.
We suggest that even the direct observation of ponderomotively squeezed states in optomechanical experiments can act as a probe of spontaneous decoherence processes on the mechanical motion.
If the mirror motion spontaneously decoheres during the interaction time of light and mirror, the ponderomotively squeezed state becomes a mixture of several states and its purity degrades.
The quantum correlation between the reflected light and the mirror motion is always effected by several decoherence mechanisms from the local environment. 
If a spontaneous decoherence mechanism exists on top, it might be observable and quantified if reference measurements can be used to quantify all environmental decoherence mechanisms.
This way, the LIGO experiment \cite{Yu2020} can in principle be used to search for unknown spontaneous (gravitational) decoherence mechanisms. 
Whether the conventional decoherence mechanisms can be quantified with a precision high enough to actually challenge well-motivated spontaneous decoherence mechanisms is not investigated here. 
However, in Ref.~\cite{Diosi2015} it is argued that even with mixed states it might be possible to observe signatures of gravitational decoherence.
In any case, the result in \cite{Yu2020} can be used to set an upper bound for spontaneous decoherence mechanisms, simply because the quantum noise could be inferred to be below the SQL.

\section{Summary and Conclusions}\label{sec:conclusions}
Gravitational-wave astronomy requires unprecedented sensitivities for measuring the tiny space-time oscillations at audio-band frequencies and below.
40\,kg mirrors that are suspended as pendulums act as space-time test masses, and light fields of several 100\,kW measure the changes in their distances.
The high mass minimises the mirrors' quantum uncertainties in position and momentum on absolute scales, the effect of the light's radiation pressure uncertainty, as well as disturbances on the mirror motion due to the environment. Mechanical resonances have high quality factors and are designed to keep away thermally driven vibrations from the relevant spectrum \cite{Rowan2005}.
The mirrors' triple pendulum suspensions are complemented by a series of passive and active seismic isolations \cite{Cumming2012,vanHeijningen2019}. Mirror surfaces are super-polished to minimise light scattering and feedback of back-scattered light that carry modulations due to the movements of the environment \cite{Vahlbruch2007,Ottaway2012}. 
The light power is high to maximise the optical GW signal on the output beam with respect to the latter's quantum uncertainty. 
Despite the high power, neither the amplitude nor the frequency of the input light carry relevant disturbances from the environment, which is realised by a a large number of passive and active laser stabilisation units \cite{Kwee2012}. 
Due to all these efforts, the light's radiation pressure produces observable correlations of the optical and mechanical quantum uncertainties.\\
A recent observation in Virgo showed that the quantum uncertainties of the light fields in the arms produce such a large differential quantum radiation pressure noise that it contributed significantly to the observatories sensitivity between 30\,Hz and 70\,Hz \cite{Acernese2020}. During this observation, Virgo used its squeeze laser to squeeze the shot noise on the output photo diode. Consequently, as described by Heisenberg's uncertainty relation, the differential quantum radiation pressure in the arms had to increase.\\
LIGO used its squeeze laser to demonstrate the effect of ponderomotive squeezing, which is additional, superimposed squeezing due to the coupling of the light's radiation pressure uncertainty and the momentum/position uncertainty of the mirror motion \cite{Yu2020}. Ponderomotive squeezing is only produced if the quantum uncertainties of reflected light and mirror motion are quantum correlated. The injected squeezing plus the ponderomotive squeezing resulted in a quantum noise that was below the standard quantum limit (SQL) between 30\,Hz and 50\,Hz. This observation was possible after classical noise of about 1.5-times higher standard deviation was subtracted. 

The quantum uncertainties of the test mass motion in LIGO and Virgo, as they were imprinted on the reflected light in \cite{Acernese2020} and \cite{Yu2020}, 
had magnitudes as expected. If they had been significantly weaker, potentially not visible at all, they would have pointed to an unknown decoherence process acting on the mirror motion. 
There are indeed hypothetical \emph{spontaneous} decoherence mechanisms, which were proposed to explain why quantum coherent effects are not observed on macroscopic objects. 
Prominent examples are `gravitationally induced spontaneous localisation' according to the Di\'{o}si-Penrose models \cite{Karolyhazy1966,Diosi1987,Diosi1989,Penrose1996,Penrose1998,Bassi2017}. 
Their typical rationals, however, lead to weak decoherence rates, which are not testable with state of the art technology \cite{Bose1999,Bernad2006, Adler2007,Kleckner2008a,Miao2010}. 
Only a modified version of gravity decoherence as conjectured in \cite{Miao2010} leads to sufficiently short decoherence times whose measurement is feasible. 
Other continuous spontaneous localisation (CSL) models that are independent of gravity were also proposed~\cite{Ghirardi1990,bassi2003dynamical,Bassi2013}.
Some of them might result in measurable decoherence effects.

Historically, optomechanical \emph{non-}Gaussian quantum states were first proposed for testing spontaneous decoherence/localisation models. In particular Schr\"odinger cat states, whose superimposed position states are macroscopically distinct and separated by more than the size of the gravitating body, seemed promising for testing gravitational decoherence models \cite{Diosi1987,Diosi1989,Ghirardi1990,Penrose1996,Bose1999,Marshall2003}. 
But these states are way beyond state of the art technology \cite{Bose1999}. Feasible are only non-Gaussian states of mechanical oscillators, whose dimension is much larger than the size of the quantum uncertainty \cite{Marshall2003,Khalili2010,Hong2017,Riedinger2018,Marinkovic2018}.

Taking this for granted, Gaussian nonclassical optomechanical states seem to us as suitable for testing spontaneous localisation models as the non-Gaussian ones. We suggest using the ponderomotively squeezed optical states. In the ideal case of zero decoherence on the light field and on the mirror motion, ponderomotively squeezed states are pure and have minimal quantum uncertainties. 
Spontaneous as well as environmental decoherence effects would be measured on a stationary system, in which continuous-wave light continuously places the optomechanical system of mirror motion and reflected light in a quantum correlated state. The various decoherence mechanisms continuously mix the state with uncorrelated thermal mechanical and optical states, and the continuous sampling on the steady-state light/mirror system
would result mixed optical states with an above minimum uncertainty product.  
Quantum tomography on ponderomotively squeezed laser fields could thus serve as tests of gravitationally induced spontaneous localisation.
Quantum tomography on the output light in GW observatories is possible by replacing the traditional photo diode by two such photo diodes in balanced homodyne detector arrangement, which needs to subsequently measure the field variance at some sideband frequency near the SQL in the most squeezed quadrature and in the one 90$^\circ$ off. Alternatively, two balanced homodyne detectors simultaneously measure the two quadrature fields, which also provides the full quantum information but additionally allows for vetoing non-stationary disturbances due to back-scattered light \cite{SteinlechnerS2013}.  
The sensitivity of searches for such unknown decoherence processes in GW observatories, however, is very low. 
The quantum correlations between the optical field and mirror movement are observable but obscured by thermal energy and other noise sources.\\

\begin{acknowledgments}
This work was supported by the Deutsche Forschungsgemeinschaft (DFG) under Germany's Excellence Strategy EXC 2121 `Quantum Universe' -- 390833306.
\end{acknowledgments}

\subsection*{Data Availability Statement}
The data that support the findings of this study are available from the corresponding author upon reasonable request.

\subsection*{Conflict of interest}
The authors have no conflicts to disclose.

\section*{References}


\end{document}